\documentclass[journal]{IEEEtran}
\usepackage{cite}

\ifCLASSINFOpdf
   \usepackage[pdftex]{graphicx}
\else
   \usepackage[dvips]{graphicx}
\fi
\usepackage{amsmath}
\usepackage{url}

\usepackage{multirow}
\usepackage{enumitem}
\usepackage{pgfplots}
\usepgfplotslibrary{statistics}

\begin{document}
%
\title{Legio: Fault Resiliency for Embarrassingly Parallel MPI Applications}
%
%
%

\author{Roberto~Rocco, 
        Davide~Gadioli,
        and~Gianluca~Palermo
\thanks{R. Rocco, D. Gadioli and G. Palermo are with the Dipartimento di Elettronica, Informazione e Bioingegneria (DEIB), Politecnico di Milano, Milan, 20133 Italy}}
\maketitle

\begin{abstract}
Due to the increasing size of HPC machines, the fault presence is becoming an eventuality that applications must face. Natively, MPI provides no support for the execution past the detection of a fault
, and this is becoming more and more constraining. With the introduction of ULFM (User Level Fault Mitigation library), it has been provided with a possible way to overtake a fault during the application execution at the cost of code modifications. ULFM is intrusive in the application and requires also a deep understanding of its recovery procedures. 

In this paper we propose Legio, a framework that lowers the complexity of introducing resiliency in an embarrassingly parallel MPI application. By hiding ULFM behind the MPI calls, the library is capable to expose resiliency features to the application in a transparent manner thus removing any integration effort. 
Upon fault, the failed nodes are discarded and the execution continues only with the non-failed ones. A hierarchical implementation of the solution has been also proposed to reduce the overhead of the repair process when scaling towards a large number of nodes.  

We evaluated our solutions on the Marconi100 cluster at CINECA, showing that the overhead introduced by the library is negligible and it does not limit the scalability properties of MPI. Moreover, we also integrated the solution in real-world applications to further prove its robustness by injecting faults. 

\end{abstract}


%
\IEEEpeerreviewmaketitle

\section{Introduction}
%
%
%
%
\IEEEPARstart{T}{he} high demands of computational science applications are leading the evolution of the current high-performance systems. This increased the complexity of HPC systems to satisfy the need for more performance. As a result, the computation capabilities are growing and will reach the exascale performances (10\textsuperscript{18} FLOPS) in the next years \cite{dongarra2009international,amarasinghe2009exascale}. This evolution is introducing new challenges in the field since problems that were overlooked before are now limiting the performance of the systems. Among these problems, there is system reliability.

Modern HPC architectures are featuring millions of cores and components, and the probability that at least one of them is the victim of a fault rises with these numbers. The mean time between failures of current systems is measured in days \cite{zheng2012scalable}, and probably in the future systems will be measured in minutes \cite{dongarra2011international}. 
With this high frequency of faults, the MTBF of the system can be lower than the application run-time. This implies that applications must feature some sort of reliability management to reach the end of the execution. 
Without any explicit management, an application would have to be restarted several times up to when it is capable to reach the end of the computation without any problem. Most applications based on MPI \cite{clarke1994mpi}, the de-facto standard for intra-process communication, lack reliability management since the standard assumes that the application executes in a controlled environment, where all the system components work properly. This problem has been solved mainly by leveraging Checkpoint-and-Restart (C/R) techniques, but with the reduction of the MTBF new solutions are needed, because the time needed for the checkpoint can easily exceed the MTBF value \cite{snir2014addressing}.

To avoid relying purely on C/R, during the years many publications developed several MPI implementations featuring reliability methodologies, such as MPICH-V \cite{bouteiller2006mpich}, rMPI \cite{ferreira2011rmpi}, or FT-MPI \cite{fagg2000ft}. These efforts try to introduce reliability methodologies directly in MPI, creating new functionalities in the existing standard. While remarkable, they received only limited support and did not solve entirely and efficiently the problem. The last effort among those is the User-Level Fault Mitigation (ULFM) \cite{bland2013post} library: it’s a collection of functions that allow the user to repair and continue its MPI execution. This work is receiving a lot of attention, mainly due to the focus on the integration in the MPI standard: the next version of MPI (4.0) will focus on reliability, and ULFM is one of the candidates to be introduced in the standard.

Various efforts (such as Fenix \cite{gamell2014exploring}, CPPC \cite{losada2017resilient}, LFLR \cite{teranishi2014toward}) have been developed on top of ULFM. This comes from the fact that ULFM by itself does not specify a way to recover the execution, it just provides methods for handling the fault and repair the involved structures. These frameworks couple ULFM with a method to restore the execution (typically C/R based) and create an all-in-one tool improving the reliability of an MPI application. 

While these frameworks enhanced the reliability of an MPI application, their usage is not transparent and the application code has to be adapted accordingly. This solution is acceptable when designing a new application, but becomes problematic when targeting an already developed one. 

This aspect is limiting the impact of those frameworks and led us towards the development of a solution that does not need changes in the application code. In this work, we limit our attention to embarrassingly parallel MPI application, a very common and scalable type of parallel program that reduces to the minimum the interactions between the processes. Embarrassingly parallel application are also envisioned to be among the first ones capable to fully exploit an exascale system.
Typically, they use MPI I/O to maximize the data transfer between computation nodes and file system, while avoiding as much as possible explicit synchronization between them.


In this paper we present Legio\footnote{The name Legio comes from the Latin word that represents a military unit of the Roman army. The name was inspired by the fact that soldiers will keep fighting even after some of their fellows perish: analogously, the library aims to make MPI processes continue their execution, despite the failures of some of them.}, a framework that introduces fault resiliency in embarrassingly parallel MPI applications. It shares many aspects with the previous frameworks, such as the usage of ULFM, but focuses more on the transparency of the integration. Since embarrassingly parallel applications can continue their execution even if some processes do not provide any result, we opted for fault resiliency. Upon noticing an error, the failed processes are discarded and the execution continues only with the non-failed ones. This approach is also faster compared to the standard C/R proposed in the other frameworks, but impacts the correctness of the application result: an acceptable trade-off for applications producing an approximate result, like for example Monte Carlo solvers \cite{pauli2015intrinsic}, or high-throughput in-silico virtual screening applications \cite{Exscalate4CoVwebsite}.

Legio supports most used MPI calls in embarrassingly parallel applications together with one-sided communication and file support, features not yet included in ULFM. We also provide an alternative solution, capable of constructing a networking layer transparent to the application to reduce the impact of a fault to a few processes. We evaluate Legio on the Marconi100 cluster at CINECA \cite{m100} to measure the introduced overhead. 
Those analyses demonstrated that the proposed framework introduces fault resiliency with only a very limited impact on the performance of the application.
%

To summarize, the contributions of this paper are the following:
\begin{itemize}
\item We propose the Legio framework able to transparently introduce fault resiliency in embarrassingly parallel applications;
\item We implemented an alternative organization of MPI communicators to improve scalability;
\item We experimentally evaluate the overheads and performance impact of the proposed solutions considering both the single MPI calls and full applications;
\end{itemize}


The remainder of the paper is organized as follows. Section~\ref{sec:background} analyzes the previous works that tried to solve the problem and introduces some definitions and knowledge useful for the following sections. Section~\ref{sec:preliminary} covers the initial exploration of the ULFM behaviour in presence of faults. Section~\ref{sec:legio} exposes the design process of the Legio framework. Section~\ref{sec:hierarchical} analyzes the hierarchical alternative of the framework. Section~\ref{sec:experiment} goes through the experimental evaluation of our work by showing the overhead at the MPI call and application-level. Section~\ref{sec:future} discuses some potential improvements of the produced implementations. Lastly, Section~\ref{sec:conclusion} concludes the paper.

\section{Background and related work}\label{sec:background}
ULFM \cite{bland2013post} can be listed among one of the most relevant efforts in the field. It specifies a set of functions to enable fault tolerance in MPI applications. The main ULFM features that we use in our approach are the following: (a) the possibility to set a communicator as out of order (revoked), (b) the possibility to remove failed processes from a communicator and obtaining a working one, (c) the possibility to agree on a result even in presence of faults, and (d) the possibility to identify failed processes.

Many frameworks have been built on top of ULFM functionalities by adding different recovery strategies. 
In particular, the integration of a C/R framework with ULFM provides an all-in-one framework to manage the insurgence of faults in a generic MPI application \cite{gamell2014exploring,losada2017resilient,shahzad2018craft,teranishi2014toward,gamell2015local}. 
These solutions opted for the recovery of a consistent state: by loading a previous checkpoint, the execution restarts from a valid point. They usually provide a simple interface to the user but require changes in the application code. While obtaining a similar result to our proposed solutions, these frameworks do not pursue transparency and, rather than opting for fault resiliency, they recreate the failed processes. These characteristics provide the possibility to create a more flexible solution, working with any MPI application, at the cost of application intrusiveness and performance penalty.



A completely different perspective is the one presented when applying algorithm-based fault tolerance \cite{du2012algorithm}, which exploits the possibility to re-compute the data of a failed process using the information of the others. 
This solution is very application-specific since it leverages data redundancy to implement a resilient method with reduced overhead. Examples are shown in the context of matrix-multiplication and LU factorization kernels, but cannot be taken into consideration for a generic MPI program.
In particular, ABFT should not be exploitable in embarrassingly parallel applications, like the one we are targeting with Legio, due to the high data independence across the processes.

A method tackling transient fault has been presented in SLIM (session layer intermediary) \cite{kalim2017non}. The solution reduces the impact of transient faults by repeating the operations. Despite SLIM works for any MPI application, it cannot be considered a valid solution in case of permanent faults. 

An effort that shares many concepts with the approach we are proposing has been presented in \cite{pauli2015intrinsic}. It uses the functionalities introduced by ULFM to manage the presence of faults in a Monte Carlo application, a typical embarrassingly parallel MPI application. The authors implemented the resiliency by removing the faulty processes from the execution and continuing only with the non-failed ones. The concept behind this solution is similar to the one proposed in this paper. However, it has been achieved by directly modifying the application code since the focus of the authors was on a specific application. With Legio, we are proposing to generalize this approach by implementing a transparent framework capable to tackle all the embarrassingly parallel applications. 



\section{Preliminary analyses} \label{sec:preliminary}
In this section, we will discuss some issues of the ULFM implementation of the MPI standard in presence of faults \cite{thakur2007open,gamell2015local}. 
Before proceeding with the analysis, we want to provide some definitions of key terms for the remaining part of the paper:
\begin{itemize}
\item A \emph{process notices a fault} when it receives the error code \texttt{MPIX\_ERR\_PROC\_FAILED} after an MPI call;
\item A \emph{faulty communicator} is a communicator in which at least a participating process is failed, but no process noticed it yet;
\item A \emph{failed communicator} is a communicator in which (at least) a participating process noticed the presence of a fault;
\end{itemize}

Using these definitions, we sum up our considerations on the MPI standard in points to better refer to them in the next sections. 
\begin{enumerate}[label=\textbf{P.\arabic*}]
\item \label{Legio:p1} Some MPI functions work in faulty and failed communicators. Some remarkable functions that expose this behaviour are \texttt{MPI\_Comm\_rank} and \texttt{MPI\_Comm\_size}, but also many operations that deal with \texttt{MPI\_Group}s. These operations are labelled as local in the MPI standard and do not require communication to complete successfully.
\item \label{Legio:p2} Point-to-point communication works in a faulty communicator, as long as the processes involved in it are not failed. They do not work in a failed communicator. 
\item \label{Legio:p3} Collective communications will not work in a failed communicator but may partially work in a faulty communicator. This behaviour comes from the fact that only some of the processes may notice the fault, while the others can complete without problems. In particular, the \texttt{MPI\_Bcast} operation exposes this behaviour, unlike \texttt{MPI\_Reduce}, \texttt{MPI\_Barrier}, and \texttt{MPI\_AllReduce}. This behaviour will be called the "Broadcast Notification Problem" (\textbf{BNP}) from now on.
\item \label{Legio:p4} File and remote memory access operations are not supported by ULFM and are likely to fail in a faulty environment (rather than raising an error, they throw a segmentation fault making the execution impossible to recover).
\item \label{Legio:p5} Communicator management functions like \texttt{MPI\_Comm\_dup} or \texttt{MPI\_Comm\_split} will not work in a faulty communicator. This includes also all the Inter-communicator related ones.
\end{enumerate}

These points are used in the next Sections to justify some of the choices done while designing the proposed framework.

\section{The Legio framework design and architecture}\label{sec:legio}
The basic idea behind the Legio framework is that it has to provide fault resiliency functionalities in embarrassing parallel applications without code intrusiveness. The proposed solution consists of the substitution of the MPI structures used (and created) by the application with others managed by Legio. This way, when a fault happens, it affects only the Legio structures, making the repair process easier and controllable by the framework. 
The MPI structures that are involved in the Legio repair process are communicators, windows, and files.

To achieve our purpose, we designed a library that behaves like an intermediary between the application and the MPI implementation. To achieve that position we exploit the profiling interface provided by the MPI standard (\texttt{PMPI}), which allows us to 
intercept every MPI call made in the parallel program. Originally thought for profiling, it can be used to inject code of different types around the target MPI call. In our work, we used \texttt{PMPI} to introduce fault resiliency using ad-hoc code and ULFM methods.


The structure substitution introduces many problems that must be addressed, all referring to the possible differences between the original and the substitute. For what concerns communicator substitution, the ranks of the processes may raise some problems: the application is expecting its rank not to change during the execution, but we may have to change the communicator due to faults and, as a consequence, ranks. Our solution must be able to transparently map ranks from the original structure to the substitute one.
Therefore, when
a failed process is involved in the communication, either by being the root of a collective call or by participating in a point-to-point operation, there are two possible courses of action.
We can ignore the failure, for example when the failed process was gathering data from the others, or we can stop the application execution, for example when the failed process is spreading important data.
The choice is done at compile-time and we provided ways to the user to configure this behaviour to better fit the application.

The presence of a fault is checked after the execution of the operation with the substitute structures: if it is confirmed, then the structures must be repaired and the operation must be repeated. Since ULFM supports communicator repair only if all the processes participate in the procedure, the error checking routine is not performed in non-collective calls. The error checking routine suffers from the \textbf{BNP} (property \ref{Legio:p3}): to avoid the problem we perform an agreement operation that combines the results obtained by all the processes into a single one equal for all.

While communicators management is enough to support many MPI functions, there are many more that do not base on them. All the operations referred to in property~\ref{Legio:p1} are left unchanged, while others need some additional structures. File operations and one-sided communication ones, in particular, leverage other structures not yet supported by ULFM so any fault may cause the program to behave indefinitely (property~\ref{Legio:p4}). Any operation that uses one of these structures must be sure of the absence of faults because we cannot repair those structures and the execution would stop. The solution adopted up to now faces the problem of having to ensure that the substitute structure is fault-free before executing the operation. To achieve this requirement, we added a call to a barrier operation before the actual function: this way the eventual presence of a fault will be recognised by the barrier and it will be possible to proceed with the repair.

These solutions allow us to support most of the MPI calls, but are not sufficient for some others: common functions like the gather and scatter operations rely on the value of the rank to provide correct behaviour, and simply running them on a substitute communicator would produce a wrong result. We decided to implement those functions as a combination of others that do not suffer from the same problem.

By following these concepts, we managed to create an implementation of our library that features support for many MPI operations\footnote{The source code of the Legio framework can be found here: \url{https://github.com/Robyroc/Legio}.}. This solution is transparent: the application needs no strict code change to support the library because it is integrated only in the linking phase.
However, the application developers must be aware that an MPI operation may be skipped, due to the rank translation problem.
Therefore, they must perform all the operations required to avoid undefined behaviour, such as buffer initialization.
We evaluated our solution to measure the overhead it introduces and its ability to handle faults: we will present the results in a later section.

\section{The Hierarchical Extension}\label{sec:hierarchical}
The ULFM standard requires that all the repair procedures involve all the processes, limiting the development of local recovery solutions where each process could repair itself independently \cite{gamell2014exploring, gamell2015local}.
This is a well-known issue, analyzed by the same authors, that leads to worse than linear scaling when we increase the number of nodes involved in the computation.
Given the modern trend to increase the experiment size, the impact of this limit increases as well.

\begin{figure}[!t]
\centering
\includegraphics[width=0.7\columnwidth]{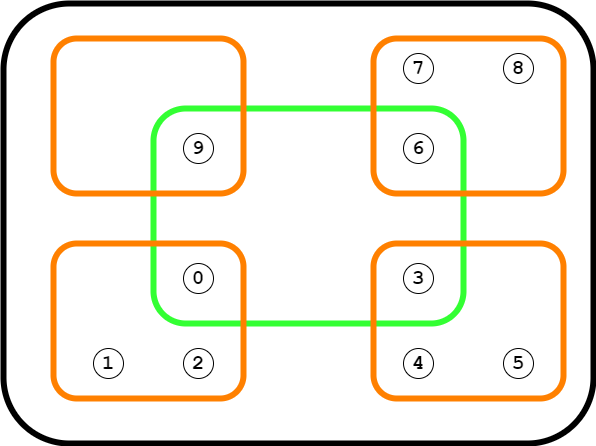}
\caption{An abstraction of a MPI application. Processes are depicted as small circles containing their rank in the target communicator. Each rounded square represents a communicator. The black one is the target communicator. The orange ones are the \textit{local\_comm}s, while the green one is the \textit{global\_comm}.}
\label{fig:topoez}
\end{figure}

To solve this issue, we propose an alternative and novel solution that avoids the \texttt{MPIX\_Comm\_shrink} usage on the entire communicator.
In particular, we developed a hierarchical approach.
At first, we split the target communicator into a set of disjoint sub-communicators (\textit{local\_comm}s).
Then, we create a new communicator (\textit{global\_comm}) that contains one process (named \textit{master}) per sub-communicator.
The \textit{master} process of a sub-communicator is the one with the lowest rank.
Figure~\ref{fig:topoez} shows the topology of the hierarchical approach.

This solution has some major properties: \textit{(a)} the number of communicators created scales linearly with the number of processes; \textit{(b)} each process can reach anyone else in the network (if not directly, via forwarding), and \textit{(c)} there is only one path from a process to another one that crosses the minimum amount of nodes.
The new communicator resembles a star topology, avoiding any communication across different \textit{local\_comm}s outside the \textit{global\_comm}.
On the main hand, this feature reduces the impact of a fault: only the processes directly communicating with the failed one will have to participate in the recovery, while the others can continue their execution seamlessly.
On the other hand, it complicates the repair procedure which depends on the role of the process.

When the faulty node is not a \textit{master}, then the repair procedure is bounded within the \textit{local\_comm}.
Otherwise, the framework needs to assign the \textit{master} role to a new process and include it in the \textit{global\_comm}.
In particular, every time the user creates a communicator (of size \(s\)), Legio creates \textit{local\_comm}s of max size \(k\).
The framework assigns each process to a \textit{local\_comm} according to its rank \(r\) i.e. a process will be assigned to the \(i\)-th \textit{local\_comm} (\textit{local\_comm\_i}) if \(i = r/k\).
Moreover, we define \textit{local\_comm\_(i+1)} as the successor of \textit{local\_comm\_(i)}, while we define \textit{local\_comm\_(i-1)} as its predecessor.
We consider the last \textit{local\_comm} the predecessor of the first \textit{local\_comm}.
The assignment of a process to a \textit{local\_comm} is final.

\begin{figure}[!t]
\centering
\includegraphics[width=0.7\columnwidth]{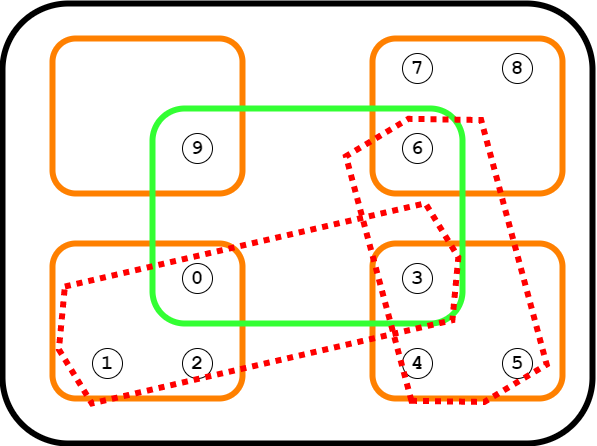}
\caption{Example of \textit{POV}s communicators in the MPI application depicted in Figure \ref{fig:topoez}. Each POV is represented as a dashed exagon. For simplicity we depict only the \textit{POV} 
of which the process with rank \(3\) is part.}
\label{fig:topopov}
\end{figure}

Due to property~\ref{Legio:p5}, when Legio manipulates a communicator, it must be fault-free.
Therefore, the framework needs to create additional communicators to complete the repair procedure, named \textit{POV}s (short for Partially OVerlapped).
Each \textit{POV} includes all the processes of a \textit{local\_comm} and the \textit{master} of the successor.
Thus, Legio creates a \textit{POV} for each \textit{local\_comm}.
Legio uses these communicators only for the repairing procedure.
Figure~\ref{fig:topopov} highlights two \textit{POV} communicators in the example depicted in Figure~\ref{fig:topoez}.

\begin{figure}[!t]
\centering
\includegraphics[width=\columnwidth]{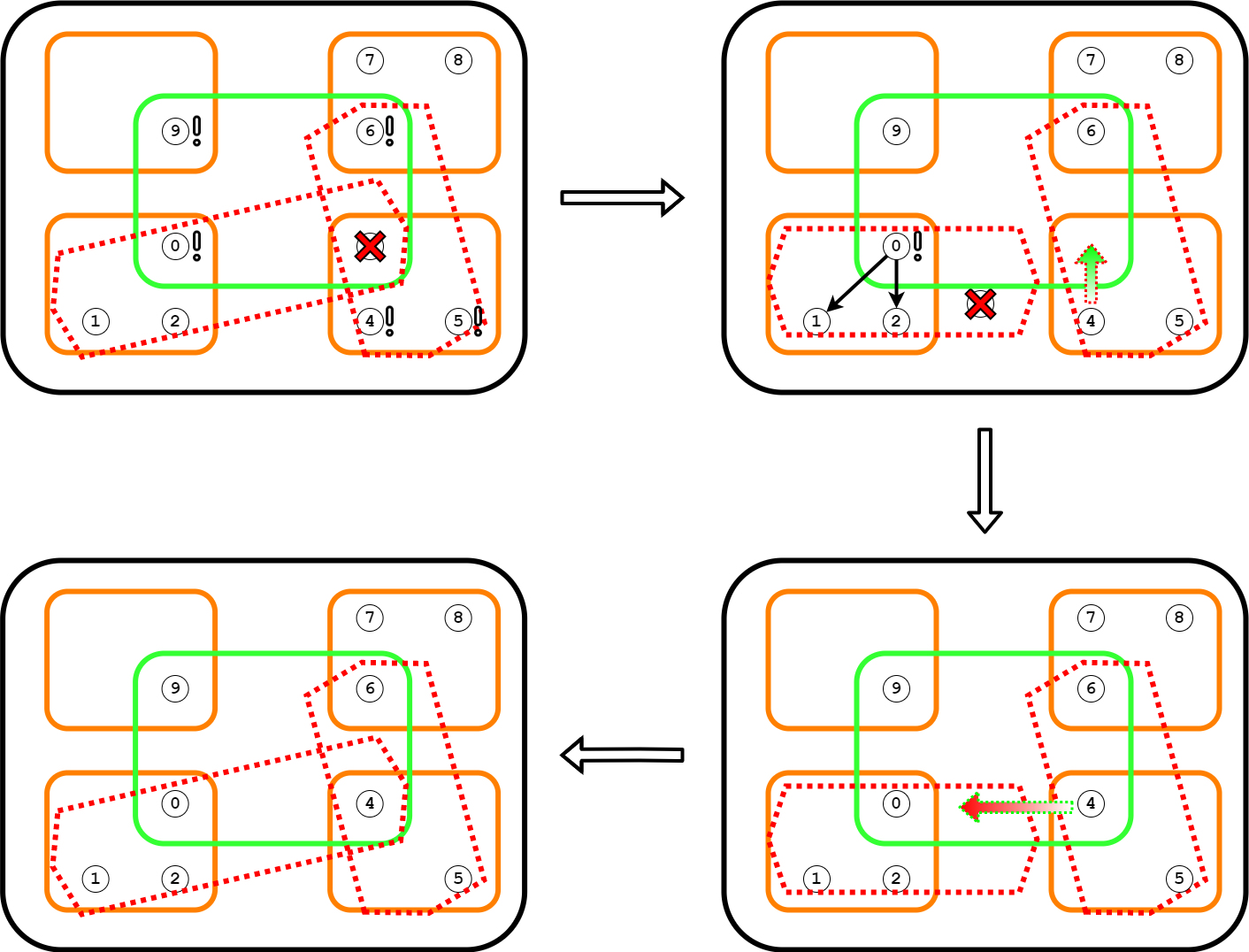}
\caption{Overview of the repair procedure when a \textit{master} fails. The communicators and processes follows the notation rules of the previous images. The red cross highlights the failed node. The exclamation marks highlight the nodes that notice the failure.
The arrows that originate from a process represent the inclusion of the process in a communicator. The arrow color represents the target communicator. The arrow border color represents the communicator used to perform the operation.
The slim black arrow represents the propagation of the failure notification.
%
}
\label{fig:procedure}
\end{figure}

Figure~\ref{fig:procedure} summarizes the required steps when we repair a failure on a \textit{master}.
The failure is noticed only by the processes in its \textit{local\_comm} and by the ones in the \textit{global\_comm}.
However, the failed process belongs to four different communicators and all of them must exclude the failed process to proceed.
The \textit{local\_comm}, its \textit{POV}, and the \textit{global\_comm} can shrink to exclude the failed \textit{master} process.
However, the \textit{master} of the predecessor needs to notify the processes in its \textit{POV} before shrinking, since they were unable to notice it directly.
In this phase of the repair procedure, the processes in the \textit{local\_comm} of the failed \textit{master} can communicate with the other processes only by using their \textit{POV} through the \textit{master} of the successor.
Legio uses this connection to include the new \textit{master} node, i.e. the process with lower rank among the ones in the \textit{local\_comm} of the failed \textit{master}, to the \textit{global\_comm}.
Then, it can use the \textit{global\_comm} to update also the predecessor POV.

Even if this procedure is composed of several steps, it reduces the cost of the repair operations because it lowers its complexity.
If we refer to \(S(x)\) as the computational cost of the shrinking operation over \(x\) processes, we can define the shrink complexity as follows:



\begin{equation}
R_H (s,k)= 
\begin{cases}
	S(k) + 2S(k + 1) + S(s/k) & \mbox{if failed master}\\
	S(k) & \mbox{otherwise}
\end{cases}
\label{eq:hcost}
\end{equation}
where \(s\) is the size of the entire communicator and we assume for simplicity that it is multiple of the maximum size of the \textit{local\_comm}s \(k\).
For the master fault case, the three terms refer to the shrinking of the \textit{local\_comm} (\(S(k)\)), the two \textit{POV}s (\(2S(k+1)\)), and the \textit{global\_comm} (\(S(s/k)\)).
The complexity depends on the role of the process, as described previously, and on the value \(s\).
When \(s\) increases, the complexity of the hierarchical approach improves with respect to \(S(s)\), i.e. shrinking the entire communicator.
In particular, there might be a minimum value of \(s\) such that the hierarchical approach will be less expensive than the normal one (for some value of \(k\)).
Formally:


\begin{equation}
\exists s_0(\forall s>s_0(\exists k | R_H (s,k) < S(s)))
\end{equation}

To answer this question we need the complexity of \(S\).
Even if we do not have a formal definition, the authors of Fenix \cite{gamell2014exploring,gamell2015local} have empirically estimated a more than linear complexity.
Under the assumption that all the processes have an equal probability of failure, Equation~\ref{eq:lowebound_complexity} and Equation~\ref{eq:upperbound_complexity} provide the relationship between the communicator size and the value of \(k\) that minimizes the overall repair complexity for the linear and quadratic case respectively.
The actual relationship lies between these two bounds.
\begin{equation}
\label{eq:lowebound_complexity}
s = \frac{k(k^2 -2)}{2}
\end{equation}
\begin{equation}
\label{eq:upperbound_complexity}
s=\sqrt{\frac{2k^2 (2k^2-1)}{3}}
\end{equation}





Even if we consider the linear case when \(s > 11\) the hierarchical approach has a lower complexity.
However, the split nature of the network introduces communication overheads since not all the processes are directly connected.
This forced us to rethink the way each operation is performed, eventually splitting the execution across the smaller communicators. In particular, we divided the supported operations in various classes, that share the same data movement characteristics:


\begin{figure}[!t]
\centering
\includegraphics[width=\columnwidth]{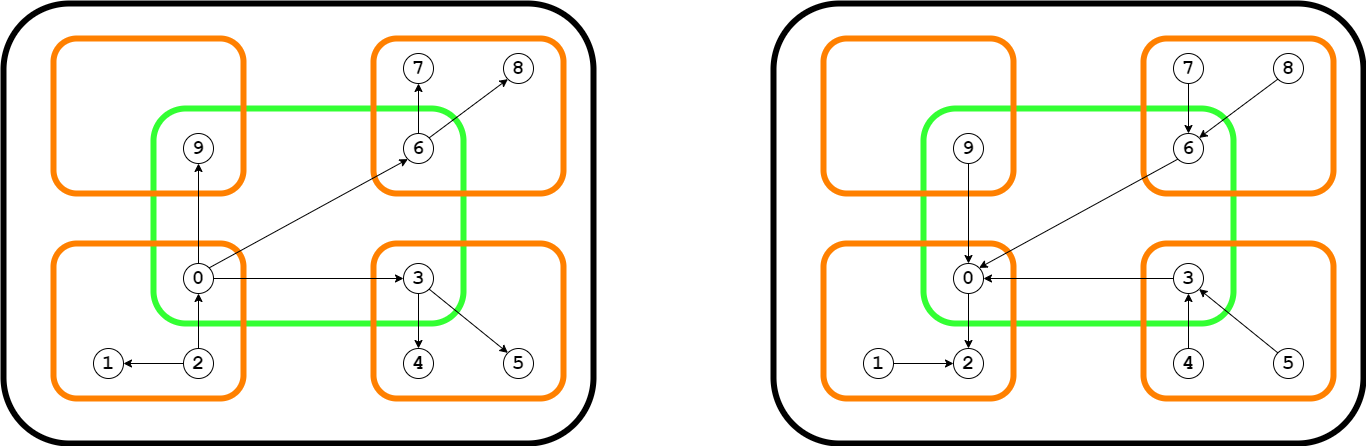}
\caption{The propagation steps in one-to-all and all-to-one operations. In both cases the root process is the one with rank \(2\).}
\label{fig:propagation}
\end{figure}

\begin{itemize}
\item \textbf{One-to-one} operations are the simplest ones since they involve only two processes. Following property~\ref{Legio:p2} and the fact that they do not need the error-checking part, we decided to run them on the entire communicator.
\item \textbf{One-to-all} operations (like \texttt{MPI\_Bcast}) involve all the processes and may cause repair. The data must go from a process to all the others, needing some sort of propagation. To execute the operation, we run it on the different parts in sequence: firstly in the \textit{local\_comm} of the root, then in the \textit{global\_comm}, and lastly in all the other \textit{local\_comm}s in parallel. Figure~\ref{fig:propagation} shows the direction of the information within the network.
\item \textbf{All-to-one} operations (like \texttt{MPI\_Reduce}) are similar to one-to-all but the data travels in the opposite direction. We followed the same propagation plan as in one-to-all but in reverse order, as shown in Figure~\ref{fig:propagation}.
\item \textbf{All-to-all} operations (like \texttt{MPI\_Allreduce}) move data from and to all the processes within the network. We decided to represent them as a combination of an all-to-one and a one-to-all operation executed sequentially.
\item \textbf{Comm-creator} operations generate new communicators. We cannot execute the operation on a \textit{local\_comm} or \textit{global\_comm} since there is the need for a unique communicator. These operations are executed on the entire communicator and may cause inefficient repairs. Nonetheless, the trade-off may be acceptable since their frequency is usually lower than the other operations.
\item \textbf{File operations} do not involve data movement between processes directly: we can use this property to make each process execute the operation on their \textit{local\_comm} without the need for any propagation mechanism.
\item \textbf{Local\_only} operations are executed by a process on its structures: no data movement is needed, so it is possible to execute the operation on the \textit{local\_comm} as done in file operations.
\end{itemize}
We decided not to support the one-sided communication functions since their implementation in a fragmented network as the one used in this hierarchical approach is not trivial.

The implementation of this solution exposes to the user two knobs: the maximum size of the \textit{local\_comm}s and a threshold value for using the hierarchical communicator.
Since this solution is an alternative to shrinking the entire communicator, we evaluated both solutions in the experimental campaign.



\section{Experimental evaluation}\label{sec:experiment}
To prove the validity of our solutions, we conducted some experiments using different benchmarks. The purpose of these experiments was to quantify and evaluate the impact of the Legio library usage on various applications. We conducted these experiments on the Marconi100 cluster at CINECA, featuring nodes with 2 x IBM POWER9 AC922 16 cores 3.1 GHz processors and 256 GB of RAM. In all the experiments done we adopted an MPI configuration featuring 32 processes per node, 1 process per physical core. 
The Legio library has been configured considering the maximum size of the \textit{local\_comm}s set to the closest optimal value following the relation obtained with the linear complexity hypothesis (Equation~\ref{eq:lowebound_complexity}).

The experiments can be divided into two groups, different for their purpose and the information they produce: the first ones involve the per-operation measurement of the overhead introduced, while the second group consists of more general applications in which we will analyze the overall impact of the library. For the first group, we used mpiBench \cite{mpiBench} to measure the overhead of the library when increasing the communication load and we used an ad-hoc code to evaluate the same parameters when increasing the network size.

\begin{figure}[!t]
\centering
\begin{tikzpicture}
\begin{axis}[
    xlabel={Packet size [B]},
    ylabel={Time [\(\mu s\)]},
    xtick=data,
    xticklabels={1k,2k,4k,8k,16k,32k,64k,128k,256k,512k,1M,2M,4M,8M,16M},
    x tick label style={rotate=45, anchor=east},
    ymode=log,
    legend pos=north west,
    ymajorgrids=true,
    grid style=dashed,
]

\addplot[
    color=blue,
    ]
    coordinates {
    (0,51.5901)(1,59.9524)(2,77.8171)(3,110.3897)(4,180.124)(5,330.6038)(6,547.4406)(7,875.8328)(8,1844.0642)(9,3167.1279)(10,2599.8997)(11,4764.8034)(12,9239.6678)(13,18239.8655)(14,34418.7941)
    };
    \addlegendentry{Legio}

\addplot[
    color=red,
    ]
    coordinates {
    (0,142.0692)(1,140.1178)(2,169.9795)(3,205.9806)(4,199.5407)(5,242.3497)(6,375.4921)(7,546.6818)(8,1029.0573)(9,2199.0139)(10,4211.2332)(11,8346.6007)(12,16606.3951)(13,33192.9681)(14,65977.0716)
    };
    \addlegendentry{Legio H}
    
\addplot[
    color=green,
    ]
    coordinates {
    (0,22.2087)(1,30.1429)(2,47.7355)(3,84.9593)(4,154.0043)(5,293.5978)(6,367.3004)(7,631.1712)(8,1129.7669)(9,2146.4767)(10,1383.1196)(11,2263.537)(12,4187.8931)(13,8068.6528)(14,13469.4742)
    };
    \addlegendentry{ULFM only}
    
\end{axis}
\end{tikzpicture}
\caption{Execution time to complete a \textit{MPI\_Reduce} by varying the message size. Each line represents a different MPI implementation.}
\label{fig:bench2}
\end{figure}
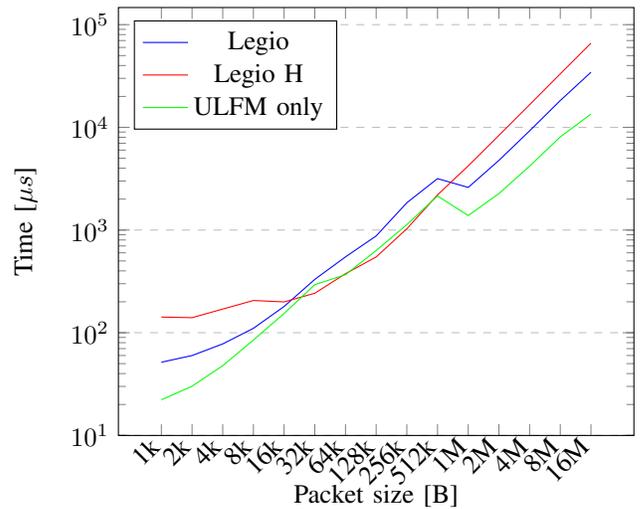

The experiments involving mpiBench were run on a 32 processes network and we analyzed the time needed to complete broadcast and reduce operations under increasing message sizes.
The mpiBench application will repeat the calls 1000 times for each message size and for each of the three versions: at first, we linked the initial Legio implementation, then the hierarchical solution, and lastly we just compiled the application with ULFM without additional libraries. Figures~\ref{fig:bench1} and~\ref{fig:bench2} show the average values of the execution times for each call.

It is possible to see how the three values share similar behaviours in terms of growths: this implies that our solutions do not damage the scalability of the MPI library with the increase of the message size.

\begin{figure}[!t]
\centering
\begin{tikzpicture}
\begin{axis}[
    xlabel={Packet size [B]},
    ylabel={Time [\(\mu s\)]},
    xtick=data,
    xticklabels={1k,2k,4k,8k,16k,32k,64k,128k,256k,512k,1M,2M,4M,8M,16M},
    x tick label style={rotate=45, anchor=east},
    ymode=log,
    legend pos=north west,
    ymajorgrids=true,
    grid style=dashed,
]

\addplot[
    color=blue,
    ]
    coordinates {
    (0,48.5127)(1,63.9169)(2,79.095)(3,100.3772)(4,139.0267)(5,212.7501)(6,359.284)(7,643.7822)(8,1151.1123)(9,1125.6722)(10,1918.5411)(11,3542.0456)(12,6699.4421)(13,9030.6244)(14,11323.8282)
    };
    \addlegendentry{Legio}

\addplot[
    color=red,
    ]
    coordinates {
    (0,138.6273)(1,162.1793)(2,177.6868)(3,239.9241)(4,327.0417)(5,490.0096)(6,820.2314)(7,1448.871)(8,2773.3312)(9,1767.9977)(10,3476.8663)(11,4081.0026)(12,4466.8775)(13,8934.2782)(14,18421.7844)
    };
    \addlegendentry{Legio H}
    
\addplot[
    color=green,
    ]
    coordinates {
    (0,22.2362)(1,34.3685)(2,41.8118)(3,67.8795)(4,93.4564)(5,149.9417)(6,274.0849)(7,504.8758)(8,997.4906)(9,1074.1139)(10,1833.8615)(11,3463.5211)(12,6628.8985)(13,8837.5025)(14,11509.2553)
    };
    \addlegendentry{ULFM only}
    
\end{axis}
\end{tikzpicture}
\caption{Execution time to complete a \textit{MPI\_Bcast} by varying the message size. Each line represents a different MPI implementation.}
\label{fig:bench1}
\end{figure}
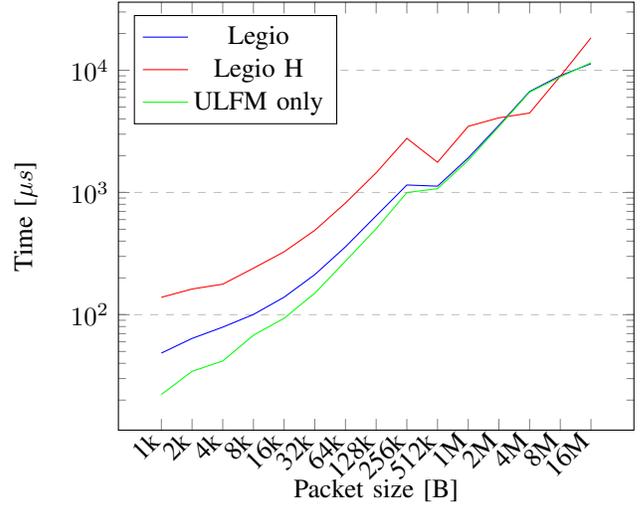

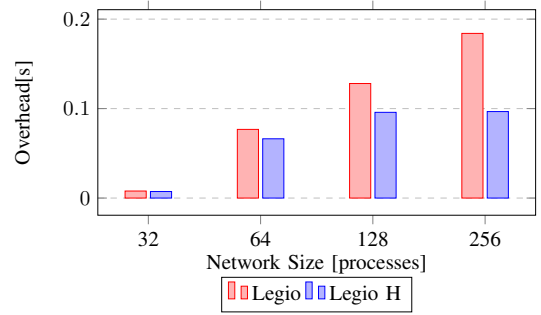
\begin{figure}[!t]
\centering
\begin{tikzpicture}[scale=0.8]
\begin{axis}[
    ymajorgrids=true,
    grid style=dashed,
    width=\columnwidth,
    height=5cm,
	xlabel={Network Size [processes]},
	ylabel={Overhead[s]},
	enlargelimits=0.15,
	ybar,
	symbolic x coords={32,64,128,256},
	xtick=data,
	legend style={at={(0.5,-0.3)},
	anchor=north,legend columns=-1},
]
\addplot[color=red, fill=red!30!white]
	coordinates {(32,0.0078587) (64,0.0767459)
		 (128,0.1280227) (256,0.1840409)};
\addplot[color=blue, fill=blue!30!white]
	coordinates {(32,0.0073136) (64,0.0662448) 
		(128,0.0958105) (256,0.0966731)};
\legend{Legio,Legio H}
\end{axis}
\end{tikzpicture}
\caption{\texttt{MPI\_Bcast} overhead by varying the network size. Each measure accumulate 100 repetitions of the operation.}
\label{fig:oh1}
\end{figure}

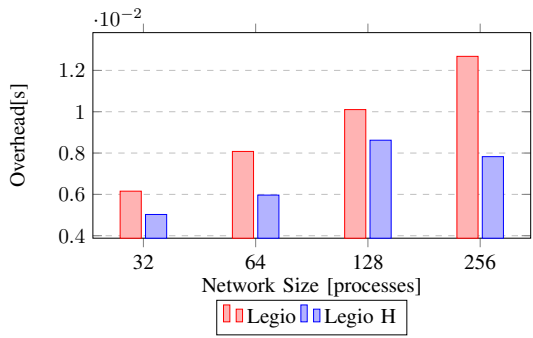
\begin{figure}[!t]
\centering
\begin{tikzpicture}[scale=0.8]
\begin{axis}[
    ymajorgrids=true,
    grid style=dashed,
    width=\columnwidth,
    height=5cm,
	xlabel={Network Size [processes]},
	ylabel={Overhead[s]},
	enlargelimits=0.15,
	ybar,
	symbolic x coords={32,64,128,256},
	xtick=data,
	legend style={at={(0.5,-0.3)},
	anchor=north,legend columns=-1},
]
\addplot[color=red, fill=red!30!white]
	coordinates {(32,0.0061553) (64,0.0080784)
		 (128,0.0101031) (256,0.0126799)};
\addplot[color=blue, fill=blue!30!white]
	coordinates {(32,0.0050274) (64,0.0059657) 
		(128,0.0086216) (256,0.0078259)};
\legend{Legio,Legio H}
\end{axis}
\end{tikzpicture}
\caption{\texttt{MPI\_Reduce} overhead by varying the network size. Each measure accumulate 100 repetitions of the operation.}
\label{fig:oh2}
\end{figure}

\begin{figure}[!t]
\centering
\begin{tikzpicture}[scale=0.8]
\begin{axis}[
    ymajorgrids=true,
    grid style=dashed,
    width=\columnwidth,
    height=5cm,
	xlabel={Network Size [processes]},
	ylabel={Overhead[s]},
	enlargelimits=0.15,
	ybar,
	symbolic x coords={32,64,128,256},
	xtick=data,
	legend style={at={(0.5,-0.3)},
	anchor=north,legend columns=-1},
]
\addplot[color=red, fill=red!30!white]
	coordinates {(32,0.0051365) (64,0.0024456)
		 (128,0.0029528) (256,0.003915)};
\addplot[color=blue, fill=blue!30!white]
	coordinates {(32,0.0097032) (64,0.0153598) 
		(128,0.0182594) (256,0.0121721)};
\legend{Legio,Legio H}
\end{axis}
\end{tikzpicture}
\caption{\texttt{MPI\_Barrier} overhead by varying the network size. Each measure accumulate 100 repetitions of the operation.}
\label{fig:oh3}
\end{figure}
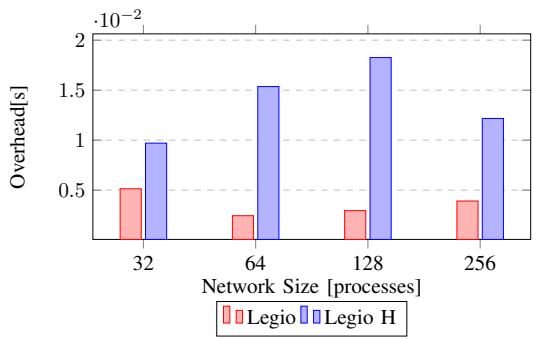

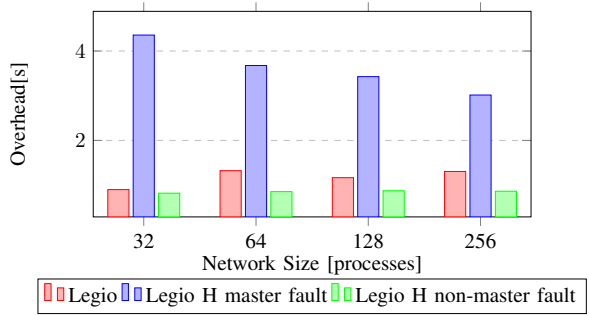
\begin{figure}[!t]
\centering
\begin{tikzpicture}[scale=0.8]
\begin{axis}[
    ymajorgrids=true,
    grid style=dashed,
    width=\columnwidth,
    height=5cm,
	xlabel={Network Size [processes]},
	ylabel={Overhead[s]},
	enlargelimits=0.15,
	ybar,
	symbolic x coords={32,64,128,256},
	xtick=data,
	legend style={at={(0.5,-0.3)},
	anchor=north,legend columns=-1},
]
\addplot[color=red, fill=red!30!white]
	coordinates {(32,0.8992943) (64,1.3230736)
		 (128,1.1665551) (256,1.3052795)};
\addplot[color=blue, fill=blue!30!white]
	coordinates {(32,4.3575465) (64,3.676376) 
		(128,3.4274439) (256,3.0159008)};
\addplot[color=green, fill=green!30!white]
	coordinates {(32,0.8194175) (64,0.8541607) 
		(128,0.8736992) (256,0.8630205)};
\legend{Legio,Legio H master fault,Legio H non-master fault}
\end{axis}
\end{tikzpicture}
\caption{Communicator repair time by varying the number of processes involved in the operation.}
\label{fig:oh4}
\end{figure}

The experiments involving the ad-hoc code have a different structure: we time each call and we compare it with the same call without the use of any Legio feature. Each call is repeated 100 times, to reduce the impact of measurement noise. Figures~\ref{fig:oh1},~\ref{fig:oh2}, and~\ref{fig:oh3} show the results obtained. We also evaluated the cost of the repair procedure by injecting a fault and completing an operation. Figure~\ref{fig:oh4} shows the results of this latter analysis: from that, it’s possible to see that the non-linearity of the shrink theorized by \cite{gamell2015local} is not present in our tests. Despite this fact, the average time to repair on a 256 core machine is lower in the hierarchical case, since the probability for a master node to fail are contained (\(1/8\)).
Using the ad-hoc code we checked also the overhead for file operations: those are more influenced by the load of the file-system rather than other aspects, so we omit those results for simplicity.

The second group contains experiments run on two embarrassingly parallel applications.
The first application is part of the NAS parallel benchmark \cite{bailey1995parallel} and it generates independent Gaussian random variates using the Marsaglia polar method.
The second one is the skeleton of a molecular docking application, which estimates the strength of the interaction between two molecules.
In this context, we have a target molecule and a database of smaller molecules that we need to evaluate to find the most promising ones.

\begin{figure}[!t]
\centering
\begin{tikzpicture}
\begin{axis}[
  boxplot/draw direction=y,
  ylabel={Time (s)},
  height=5cm,
  ymajorgrids=true,
  grid style=dashed,
  boxplot={
      %
      %
      draw position={1/4 + floor(\plotnumofactualtype/3) + 1/4*mod(\plotnumofactualtype,3)},
      %
      box extend=0.2,
  },
  x=1.5cm,
  xtick={0,1,2,...,50},
  x tick label as interval,
  xticklabels={%
      {32},%
      {64},%
      {128},%
      {256},%
  },
  xlabel={Network size [processes]},
  x tick label style={
      text width=2.5cm,
      align=center
  },
  cycle list={{red},{blue},{green}},
]

\addplot
table[row sep=\\,y index=0] {
data\\
125.546\\
125.901\\
124.593\\
129.146\\
127.925\\
124.886\\
124.931\\
125.009\\
126.409\\
126.066\\
};
\addplot
table[row sep=\\,y index=0] {
data\\
125.499\\
125.963\\
124.074\\
123.545\\
126.484\\
124.532\\
125.315\\
125.283\\
127.823\\
124.35\\
};
\addplot
table[row sep=\\,y index=0] {
data\\
124.309\\
123.255\\
132.17\\
125.565\\
129.389\\
123.318\\
124.327\\
123.606\\
126.135\\
122.339\\
};

\addplot
table[row sep=\\,y index=0] {
data\\
71.061\\
74.306\\
74.051\\
76.837\\
75.648\\
74.192\\
71.777\\
73.052\\
72.67\\
73.018\\
};
\addplot
table[row sep=\\,y index=0] {
data\\
74.829\\
79.428\\
73.143\\
72.208\\
71.929\\
70.215\\
69.958\\
72.219\\
75.86\\
72.049\\
};
\addplot
table[row sep=\\,y index=0] {
data\\
72.924\\
72.245\\
69.404\\
70.036\\
72.534\\
72.015\\
74.416\\
70.077\\
70.449\\
73.663\\
};

\addplot
table[row sep=\\,y index=0] {
data\\
43.189\\
42.343\\
45.173\\
44.019\\
43.594\\
47.826\\
48.214\\
45.981\\
51.037\\
49.685\\
};

\addplot
table[row sep=\\,y index=0] {
data\\
46.557\\
43.265\\
47.242\\
43.605\\
46.578\\
46.85\\
46.556\\
45.685\\
42.863\\
44.929\\
};

\addplot
table[row sep=\\,y index=0] {
data\\
42.133\\
43.498\\
44.676\\
42.177\\
45.31\\
45.203\\
45.063\\
42.408\\
45.95\\
44.122\\
};

\addplot
table[row sep=\\,y index=0] {
data\\
31.938\\
29.515\\
32.05\\
31.187\\
36.684\\
33.018\\
33.713\\
32.278\\
38.041\\
34.921\\
};

\addplot
table[row sep=\\,y index=0] {
data\\
30.568\\
30.631\\
33.102\\
29.636\\
30.913\\
29.296\\
30.98\\
31.183\\
32.359\\
29.881\\
};

\addplot
table[row sep=\\,y index=0] {
data\\
32.059\\
33.077\\
32.025\\
31.843\\
32.314\\
32.07\\
32.11\\
33.645\\
32.453\\
32.135\\
};

\end{axis}
\end{tikzpicture}
\caption{Execution time distribution of the EP benchmark by varying the number of processes involved and the MPI implementation.}
\label{fig:nas}
\end{figure}
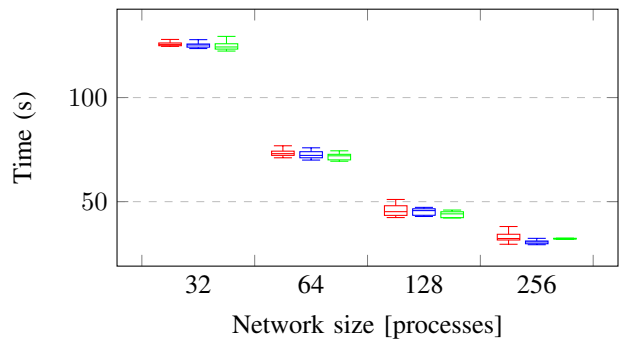

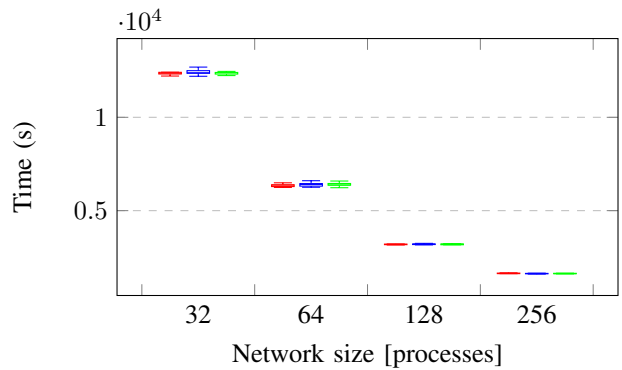
\begin{figure}[!t]
\centering
\begin{tikzpicture}
\begin{axis}[
  boxplot/draw direction=y,
  ylabel={Time (s)},
  height=5cm,
  ymajorgrids=true,
  grid style=dashed,
  boxplot={
      %
      %
      draw position={1/4 + floor(\plotnumofactualtype/3) + 1/4*mod(\plotnumofactualtype,3)},
      %
      box extend=0.2,
  },
  x=1.5cm,
  xtick={0,1,2,...,50},
  x tick label as interval,
  xticklabels={%
      {32},%
      {64},%
      {128},%
      {256},%
  },
  xlabel={Network size [processes]},
  x tick label style={
      text width=2.5cm,
      align=center
  },
  cycle list={{red},{blue},{green}},
]

\addplot
table[row sep=\\,y index=0] {
data\\
12210.13\\
12404.952\\
12329.191\\
12401.096\\
12415.531\\
12561.893\\
12581.491\\
12350.408\\
12310.645\\
12360.234\\
};
\addplot
table[row sep=\\,y index=0] {
data\\
12197.511\\
12563.556\\
12495.134\\
12501.446\\
12377.856\\
12689.631\\
12377.68\\
12385.955\\
12335.975\\
12483.942\\
};
\addplot
table[row sep=\\,y index=0] {
data\\
12234.599\\
12312.175\\
12319.02\\
12400.377\\
13070.971\\
12406.317\\
12382.082\\
12363.225\\
12383.953\\
12459.761\\
};

\addplot
table[row sep=\\,y index=0] {
data\\
6384.521\\
6294.847\\
6491.123\\
6295.067\\
6393.823\\
6329.56\\
6388.099\\
6283.402\\
6403.399\\
6243.045\\
};
\addplot
table[row sep=\\,y index=0] {
data\\
6406.404\\
6493.312\\
6253.487\\
6603.515\\
6403.707\\
6374.48\\
6467.012\\
6356.622\\
6431.765\\
6269.052\\
};
\addplot
table[row sep=\\,y index=0] {
data\\
6363.71\\
6363.174\\
6585.928\\
6440.826\\
6586.541\\
6469.284\\
6367.655\\
6433.989\\
6228.213\\
6377.795\\
};

\addplot
table[row sep=\\,y index=0] {
data\\
3197.49\\
3170.337\\
3160.012\\
3216.157\\
3203.706\\
3307.04\\
3211.267\\
3183.172\\
3189.413\\
3185.419\\
};

\addplot
table[row sep=\\,y index=0] {
data\\
3217.398\\
3165.081\\
3194.05\\
3216.792\\
3207.086\\
3227.726\\
3213.181\\
3248.077\\
3199.917\\
3173.058\\
};

\addplot
table[row sep=\\,y index=0] {
data\\
3217.625\\
3171.451\\
3168.004\\
3223.915\\
3206.945\\
3231.564\\
3209.99\\
3212.411\\
3204.499\\
3172.627\\
};

\addplot
table[row sep=\\,y index=0] {
data\\
1628.551\\
1632.09\\
1632.957\\
1625.064\\
1631.813\\
1624.624\\
1650.469\\
1660.974\\
1668.81\\
1660.095\\
};

\addplot
table[row sep=\\,y index=0] {
data\\
2091.139\\
1601.759\\
1621.921\\
1632.253\\
1629.436\\
1612.475\\
1638.082\\
1636.794\\
1639.515\\
1651.46\\
};

\addplot
table[row sep=\\,y index=0] {
data\\
1638.623\\
1611.977\\
1624.87\\
1627.597\\
1628.046\\
1614.893\\
1644.716\\
1748.542\\
1643.708\\
1650.359\\
};

\end{axis}
\end{tikzpicture}
\caption{Execution time distribution of the molecular docking application by varying the number of processes involved and the MPI implementation.}
\label{fig:focker}
\end{figure}

We use the ``C'' size workload to run the NAS application, and the measurements refer to the successive execution of 40 runs.
We use a database with 113K molecules to test the docking application. We ran both of them in various configurations in terms of the number of MPI processes and MPI implementation.
In particular, we use 32, 64, 128, and 256 processes and choose between one of our implementations or only ULFM. We repeat experiments in each configuration 10 times, extracting all the execution times. The results of these executions can be seen in Figures~\ref{fig:nas} and~\ref{fig:focker}: it is easy to see that the overhead is negligible and the usage of Legio does not impact heavily the execution times.

Those experiments validate our approach since they respect the requirements of low overhead introduction and transparency.
Moreover, both the prototypes proved effective for the embarrassingly parallel applications tested and can continue the execution in presence of faults in a manner of seconds.

\section{Discussion on Introducing the C/R feature}\label{sec:future}

The current implementation of Legio relies on the fact that the MPI application can produce useful results even if a failed process can be removed from the full set. In several embarrassing parallel applications, this is true, but there is a fraction of them that does not fulfill this hypothesis. In this direction, we already evaluated the possibility to introduce a C/R feature in the library thus obtaining the possibility to recover failed processes transparently.

As discussed in Section~\ref{sec:background}, many other efforts combined C/R frameworks with ULFM (\cite{gamell2014exploring,losada2017resilient,shahzad2018craft,teranishi2014toward,gamell2015local}). However, none of them focus towards the transparency. All these efforts base on application-level C/R frameworks which allow the application to explicitly specify what should be saved and when to do so during the check-pointing phase. 
An alternative to application-level C/R frameworks are those working at system-level. System-Level C/R frameworks implement a transparent approach at the cost of a large overhead for both check-pointing (all the system status has to be saved) and restart phase.

Legio has in transparency one of its key features, thus forcing us towards system-level C/R frameworks. However, considering the type of application we are targeting, the characteristic that we want to take out from a C/R framework is not to restart the entire application, but only the failed processes. The possibility to restart only a part of the network is not a common feature in system-level C/R frameworks. Usually, these frameworks are designed to consider the absence of fault mitigation mechanisms inside the application, so they assume that in case of fault all the processes must be restarted. Moreover, they tend not to split the checkpoint information of the various processes because they would lose significance without all the others. The restart part may also lead to problems: without knowing the details about the application, it may be difficult to load a system-level checkpoint on a process created by the application.

However, among all the efforts produced in literature, recently we found in MANA \cite{garg2019mana} a support in that direction. It provides system-level checkpointing (no application intrusiveness), the possibility to migrate processes (implying the division of the data per process), and flexibility on MPI versions upon restart. Our idea is to exploit the per-process data checkpointing offered by MANA to actually restore only the failed process. While everything seems ready for the integration, MANA is still designed for global recovery and the steps towards local recovery are part of our on-going work.

\section{Conclusion}\label{sec:conclusion}

This paper presents Legio, a framework designed to offer resiliency to embarrassing parallel MPI applications. The work makes the absence of intrusiveness in the target application one of the key elements. Indeed, the library makes use of the PMPI interface to wrap the MPI call and to implement all the required actions to manage failed processes.
ULFM has been used as a base for the implementation of Legio.

In the paper, an extension towards a hierarchical implementation has been done to reduce the overhead of the repair process in case of a large number of nodes involved. Within the document, also a theoretical analysis of when to apply the hierarchical version has been discussed.

The experimental evaluations considering both per-MPI-call and application-level evaluations demonstrate the efficiency of the implemented framework, proving how the solution can be used in embarrassingly parallel applications without affecting the overall performance. 



Overall, the problem of dealing with faults in an MPI application is still complex and needs additional efforts. The proposed solution is very specific both in the target application and the recovery policy (fault resiliency), and it does not apply to a generic MPI application. Nonetheless, embarrassingly parallel applications are intrinsically scalable and are compatible with the future exascale architectures, and in those systems, fault tolerance will be even more important.


\ifCLASSOPTIONcaptionsoff
  \newpage
\fi



%

\bibliographystyle{IEEEtran}
\bibliography{biblio}

%








\end{document}